\documentclass[12pt,a4paper]{article}
\usepackage{amsfonts,latexsym}
\usepackage{amsmath,amssymb}
\usepackage{graphicx,color}
\usepackage{multirow}
\usepackage{slashbox}
\numberwithin{equation}{section} \oddsidemargin 0 mm
\renewcommand{\thefootnote}{\fnsymbol{footnote}}
\newcommand{\nn}{\nonumber}

\begin{document}
\vspace{12mm}

\begin{center}
{{{\Large {\bf Stability of Schwarzschild black hole in f(R) gravity with the dynamical Chern-Simons term}}}}\\[10mm]

{Taeyoon Moon$^{a}$\footnote{e-mail address: tymoon@sogang.ac.kr}
and  Yun Soo Myung$^{b}$\footnote{e-mail address:
ysmyung@inje.ac.kr},
}\\[8mm]

{{${}^{a}$ Center for Quantum Space-time, Sogang University, Seoul, 121-742, Korea\\[0pt]
${}^{b}$ Institute of Basic Sciences and School of Computer Aided Science, Inje University Gimhae 621-749, Korea}\\[0pt]
}
\end{center}
\vspace{2mm}

\begin{abstract}
We perform the stability analysis of the Schwarzschild black hole in
$f(R)$ gravity with the parity-violating Chern-Simons (CS) term
coupled to a dynamical scalar field $\theta$. For this purpose, we
transform the $f(R)$ gravity into the scalar-tensor theory by
introducing a scalaron $\phi$, providing the dynamical Chern-Simons
modified gravity with two scalars. The perturbation equation for the scalar
$\theta$ is coupled to the odd-parity metric perturbation equation,
providing a system of  two coupled second order equations, while the
scalaron is coupled to the even-parity perturbation equation. This
implies that the CS coupling affects the Regge-Wheeler equation,
while $f(R)$ gravity does not affect the Zerilli equation.  It turns
out that the Schwarzschild black hole is stable against the external
perturbations if the scalaron is free from the tachyon.

\end{abstract}
\vspace{5mm}

{\footnotesize ~~~~PACS numbers: }

\vspace{1.5cm}

\hspace{11.5cm}{Typeset Using \LaTeX}
\newpage
\renewcommand{\thefootnote}{\arabic{footnote}}
\setcounter{footnote}{0}


\section{Introduction}
$f(R)$ gravities~\cite{NO,sf,NOuh} have much attention as one of
strong candidates for explaining the current accelerating
universe~\cite{SN}.  $f(R)$ gravities can be considered as Einstein
gravity  with an additional scalar (scalaron).   For example, it was
shown that the metric-$f(R)$ gravity is equivalent to the
$\omega_{\rm BD}=0$ Brans-Dicke   theory with a certain
potential~\cite{FT}.

On the other hand, the Chern-Simons (CS) modified gravity was
obtained by adding a parity-violating CS term to the
Einstein-Hilbert action, where the CS term couples to gravity via a
CS scalar field $\theta$~\cite{JP}. Originally the coupled scalar
field $\theta$ was considered as a prescribed function, but on later
this choice was not regarded as the well-motivated one. Indeed, the
dynamical Chern-Simons (DCS) modified  gravity has been formulated
by treating the scalar field $\theta$ as a dynamical field~\cite{SECK}. For a
review on the CS modified gravity, its astrophysical consequences,
see~\cite{AY} and for its critical gravity on the AdS$_4$ spaceimes, see~\cite{beato,MM}

It is very interesting to investigate the Schwarzschild black hole
obtained from {\it a modified gravity of the f(R) gravity with the
dynamical CS term} because astrophysical black holes are the most
promising objects to probe the strongly gravitational field region
of a modified gravity. The first study of the $f(R)$-black hole
stability has very recently been performed in the $f(R,G)$
gravity~\cite{FST}. In its scalar-tensor theory~\cite{MMS}, the
even-parity perturbations were affected by the scalaron and thus the
black hole was stable against the whole perturbations if the
scalaron did not have a tachyonic mass.    In the context of the DCS
modified gravity, the black hole perturbation has been carried out
in~\cite{Yunes}, which indicates  that if the background CS scalar
$\bar{\theta}$ is a non-trivial, there was a serious mixing between
odd-and even-parity metric perturbations.  On the other hand, if
$\bar{\theta}=0$ or const., odd-and even-perturbations were
decoupled as in Einstein gravity and odd-perturbations are
affected only by the CS scalar field~\cite{CG}. The odd-parity and
CS scalar perturbations were described by a coupled system of two
second-order equations, which has shown that the black hole is
stable in the DCS modified gravity~\cite{MPCG}.

Very recently, there was a perturbation study on the black hole in
the context of $f(R,C)$ modified gravity with $C$ the CS
term~\cite{MS}. The black hole is unstable because the perturbed
Hamiltonian is not bounded from below, due to the CS term. In order
to avoid the instability, either $\bar{R}={\rm const.}$ or
$\frac{\partial^2f}{\partial R\partial C}=0$ is required. In this
case, number of physically propagating degrees of freedom are three,
one from odd-parity and two from even-party and scalaron because the
$f(R,C)$ modified gravity belongs to the non-dynamical CS modified
gravity. Those modes are too strongly coupled to decouple three
independent modes, which shows a distinctive feature of a parity-violating
theory. However, the no-ghost condition of $\frac{\partial
f(R,C)}{\partial R}>0$ and no-tachyon condition of $\frac{\partial^2
f(R,C)}{\partial R^2}>0$ survive as in $f(R)$ gravities.

In this work, we wish to perform the stability analysis of the
Schwarzschild black hole in $f(R)$ gravity with the parity-violating
CS term coupled to a dynamical scalar field $\theta$.  In order to
avoid the difficulty with fourth-order derivative terms, we first
transform the $f(R)$ gravity into the scalar-tensor theory by
introducing a scalaron $\phi$. This will  provide the DCS modified
gravity with two scalars, which means that four modes are physically propagating
degrees of freedom. Interestingly, the perturbation equation for the
CS scalar $\theta$ is coupled to the odd-parity metric perturbation
equation, providing a system of two coupled second-order equations,
while the scalaron $\phi$ is coupled to the even-parity perturbation
equation. This enables us to perform the stability analysis of the
Schwarzschild black hole obtained from $f(R)$+DCS modified gravity
theory completely.
To make all things clear, we mention our notations.
 The metric signature is $(-,+,+,+)$. The Riemann, Ricci tensor
and Levi-Civita tensor are defined by
\begin{eqnarray}
R^{\rho}_{~\sigma\mu\nu}=\partial_{\mu}\Gamma^{\rho}_{\nu\sigma}
-\partial_{\nu}\Gamma^{\rho}_{\mu\sigma}+
\Gamma^{\rho}_{\mu\lambda}\Gamma^{\lambda}_{\nu\sigma}-
\Gamma^{\rho}_{\nu\lambda}\Gamma^{\lambda}_{\mu\sigma},~~~
R_{\mu\nu}=R^{\rho}_{~\mu\rho\nu},~~~
\epsilon^{tr\varphi_1\varphi_2}=\frac{1}{\sqrt{-g}}. \nonumber
\end{eqnarray}

\section{$f(R)$ gravity with the DCS term}
Let us  consider $f(R)$ gravity with the dynamical Chern-Simons
term in four dimensions which is given by
\begin{eqnarray}
S=\frac{1}{2\kappa^2}\int d^4 x\sqrt{-g} \Bigg[
f(R)+\frac{\theta}{4}{}^{*}RR
-\alpha\nabla_{\mu}\theta\nabla_{\mu}\theta\Bigg]\label{Action}
\end{eqnarray}
where $\kappa^2=8\pi G$, $\alpha$ is a dimensional constant, and
${}^{*}RR={}^{*}R^{\eta~\mu\nu}_{~\xi}R^{\xi}_{~\eta\mu\nu}$ is the
Pontryagin density with
\begin{eqnarray}
{}^{*}R^{\eta~\mu\nu}_{~\xi}=\frac{1}{2}\epsilon^{\mu\nu\rho\sigma}R^{\eta}_{~\xi\rho\sigma}.
\end{eqnarray}
Here $\epsilon^{\mu\nu\rho\sigma}$ denotes the four-dimensional
Levi-Civita tensor. It is well known that the action can be
rewritten by introducing a scalaron  field $\phi$ as follows~\cite{Olmo}:
\begin{eqnarray}
S=\frac{1}{2\kappa^2}\int d^4 x\sqrt{-g} \Bigg[ \phi
R-V(\phi)+\frac{\theta}{4}{}^{*}RR -\alpha\nabla_{\mu}\theta\nabla_{\mu}\theta\Bigg]
\end{eqnarray}
with the potential $V(\phi)=\phi A(\phi)-f(A(\phi))$. Note that  the mass dimensions of $\phi$, $\theta$, and  $\alpha$ are
given by $[\phi]=0,~[\theta]=-2,~[\alpha]=4$, respectively. Varying
for the fields $g_{\mu\nu},~\phi$, and $\theta$ lead to the
following equations:
\begin{eqnarray}
&&\phi\left(R_{\mu\nu}-\frac{1}{2}g_{\mu\nu}R\right)+\frac{1}{2}g_{\mu\nu}V(\phi)
+\Big(g_{\mu\nu}\nabla^2-\nabla_{\mu}\nabla_{\nu}\Big)\phi\nn\\
&&\hspace*{8em}=-C_{\mu\nu}
+\alpha\Big(\nabla_{\mu}\theta\nabla_{\nu}\theta
-\frac{1}{2}g_{\mu\nu}\nabla_{\rho}\theta\nabla^{\rho}\theta\Big),\label{eomg}\\
&&\hspace*{3em}R=V'(\phi),\label{eomphi}\\
&&\hspace*{3em}\nabla^2\theta=-\frac{1}{8\alpha}{}^{*}RR
\label{eomtheta}
\end{eqnarray}
where ${}^{\prime}$ denotes differentiation with respect to $\phi$,
 and $C_{\mu\nu}$ takes the form
\begin{eqnarray}\label{cotton}
C_{\mu\nu}=\nabla_{\rho}~\theta~\epsilon^{\rho\sigma
\gamma}_{~~~(\mu}\nabla_{|\gamma|}R_{\nu)\sigma}+\frac{1}{2}\nabla_{\rho}\nabla_{\sigma}
~\theta~\epsilon_{(\nu}^{~~\rho \gamma
\delta}R^{\sigma}_{~~\mu)\gamma \delta}.
\end{eqnarray}
 We take the trace  of
(\ref{eomg}) to rewrite (\ref{eomphi}) as  the scalaron equation
\begin{equation}\label{eomphi1}
3\nabla^2\phi+2V(\phi)-\phi
V'(\phi)=-2\alpha\nabla_{\mu}\theta\nabla^{\mu}\theta.
\end{equation}
Also  we can express  Eq.({\ref{eomg}) to be
\begin{eqnarray}\label{eomg1}
\phi
R_{\mu\nu}-\frac{1}{2}g_{\mu\nu}V(\phi)-\frac{1}{2}g_{\mu\nu}\nabla^2\phi
-\nabla_{\mu}\nabla_{\nu}\phi=-C_{\mu\nu}
+\alpha\nabla_{\mu}\theta\nabla_{\nu}\theta.
\end{eqnarray}
Taking the restricted background values\footnote{ When taking
these values, it gives  the background spacetimes with the
constant curvature scalar $\bar{R}$ which provides  an easy step
to find the solution of $f(R)$ gravity. In obtaining the constant
curvature-black hole solutions (for example, Schwarzschild and
Schwarzschild-(A)dS black holes), it seems that there is no
difference between $\bar{\theta}=0$ and $\bar{\theta}={\rm
const.}$.  Hence we choose $\bar{\theta}={\rm const.}$ here. Note
that $\bar{\phi}$ corresponds to $f'(\bar{R})$ in the original
$f(R)$ gravity and
$\bar{R}=V'(\bar{\phi})=2V(\bar{\phi})/\bar{\phi}$ from
Eqs.(\ref{eomphi}) and (\ref{eomphi1}).  In this work, we focus on
the  Schwarzschild black hole solution with $\bar{R}=0$, which
implies that $V'(\bar{\phi})=V(\bar{\phi})=0$. We mention that
this is possible to occur  when choosing  a limited form of $f(R)$
gravity: $f(R)=a_1R+a_2 R^2+\cdots$~\cite{PPDM,BS,myungrot}. In
this case, one finds that  $\bar{\phi}=a_1$. } as
 \begin{equation}
\bar{\theta}={\rm const.},~\phi=\bar{\phi}={\rm const.},
~V(\bar{\phi})=V'(\bar{\phi})=0,~V''(\bar{\phi})\not=0,\end{equation}
the solution to the Eqs.(\ref{eomg}), (\ref{eomphi}) and
(\ref{eomtheta}) is given by the Schwarzschild spacetime
\begin{eqnarray}
ds_{Sch}^2&=&\bar{g}_{\mu\nu}dx^{\mu}dx^{\nu}\nn\\
&=&-f(r)dt^2+\frac{dr^2}{f(r)}+r^2(d\varphi_1^2+\sin^2\varphi_1
d\varphi_2^2)
\end{eqnarray}
with the metric function \begin{equation} f(r)=1-\frac{2M}{r}. \end{equation}

 Now we introduce the perturbation
around the background metric as
\begin{eqnarray} \label{m-p}
g_{\mu\nu}=\bar{g}_{\mu\nu}+h_{\mu\nu}.
\end{eqnarray}
The perturbations around the background solution $\bar{\phi}$ and
$\bar{\theta}$ are given by
\begin{eqnarray}\label{s-p}
\theta=\bar{\theta}+\delta\theta,~~~\phi=\bar{\phi}+\delta\phi.
\end{eqnarray}
The linearized equation to (\ref{eomg1}) can be written by
\begin{eqnarray}\label{pertg}
\bar{\phi}\delta
R_{\mu\nu}(h)-\frac{1}{6}\bar{g}_{\mu\nu}\bar{\phi}V''(\bar{\phi})\delta\phi
-\bar{\nabla}_{\mu}\bar{\nabla}_{\nu}\delta{\phi}=-\delta C_{\mu\nu}
\end{eqnarray}
where the linearized quantities of $\delta R_{\mu\nu}(h),~\delta
R(h),$ and $\delta C_{\mu\nu}(h)$ take the forms
\begin{eqnarray}\label{cottonp0}
\delta
R_{\mu\nu}(h)&=&\frac{1}{2}\left(\bar{\nabla}^{\gamma}\bar{\nabla}_{\mu}
h_{\nu\gamma}+\bar{\nabla}^{\gamma}\bar{\nabla}_{\nu}
h_{\mu\gamma}-\bar{\nabla}^2h_{\mu\nu}-\bar{\nabla}_{\mu} \bar{\nabla}_{\nu} h\right)\nn\\
\label{cottonp1}\delta
R(h)&=&\bar{\nabla}^{\mu}\bar{\nabla}^{\nu}h_{\mu\nu}-\bar{\nabla}^2h
\nn\\
\label{cottonp2}\delta
C_{\mu\nu}(h)&=&\frac{1}{2}\bar{\nabla}_{\rho}\bar{\nabla}_{\sigma}
~\delta\theta~\epsilon_{(\nu}^{~~\rho \gamma
\delta}\bar{R}^{\sigma}_{~~\mu)\gamma \delta}.
\end{eqnarray}
In these expressions, the ``overbar'' denotes the background
quantities. From Eq.(\ref{eomphi1}) and (\ref{eomtheta}), we obtain
the linearized-scalaron equation
\begin{eqnarray}\label{pertphi}
\Big[\bar{\nabla}^2-\frac{1}{3}\bar{\phi}V''(\bar{\phi})\Big]\delta\phi=0
\end{eqnarray}
and the linearized-$\theta$ equation
\begin{eqnarray}\label{perttheta}
\bar{\nabla}^2\delta\theta=-\frac{1}{4\alpha}
\epsilon^{\mu\nu\rho\sigma}\bar{R}^{\eta}_{~\xi\mu\nu}
\bar{\nabla}_{\rho}\bar{\nabla}_{\eta}h^{\xi}_{\sigma}
\end{eqnarray}

\section{Perturbation analysis}

The metric perturbations $h_{\mu\nu}$ are classified according to
the transformation properties under parity, namely odd sector
($h_0,~h_1$) and even sector ($H_0,~H_1,~H_2,K$). However it is
nontrivial task to show how the decoupling process goes  with two
scalar fields ($\delta\theta,~\delta\phi$) well.\footnote{It turns
out that for $\bar{\theta}\neq {\rm const.}$, there was  mixing
between odd and even modes in Chern-Simons modified
gravity~\cite{Yunes}. Here we can avoid this difficulty by
choosing $\bar{\theta}={\rm const.}$.} In order to see this
explicitly, we  must consider the full metric perturbation as
\begin{eqnarray} h_{\mu\nu}=\left(
\begin{array}{cccc}
H_0(r)Y & H_1(r)Y &
-\frac{\partial_{\varphi_2}Y}{\sin\varphi_1}h_0(r)& \sin\varphi_1
\partial_{\varphi_1}Yh_0(r)\cr
H_1(r)Y & H_2(r)Y
&-\frac{\partial_{\varphi_2}Y}{\sin\varphi_1}h_1(r)& \sin\varphi_1
\partial_{\varphi_1}Yh_1(r) \cr
-\frac{\partial_{\varphi_2}Y}{\sin\varphi_1}h_0(r)&
-\frac{\partial_{\varphi_2}Y}{\sin\varphi_1}h_1(r) & r^2YK(r) & 0
\cr \sin\varphi_1
\partial_{\varphi_1}Yh_0(r) & \sin\varphi_1
\partial_{\varphi_1}Yh_1(r) & 0 &
r^2\sin^2\varphi_1YK(r)
\end{array}
\right) e^{-ikt} \label{pmetric}
\end{eqnarray}
with $Y\equiv Y^{LM}(\varphi_1,\varphi_2)$ spherical harmonics. The
form of $\delta\theta$ and $\delta \phi$ are given by
\begin{eqnarray}\label{perttheta1}
\delta\theta=\frac{\psi(r)}{r}Ye^{-ikt},
~~~\delta\phi=\frac{\Phi(r)}{r}Ye^{-ikt}.
\end{eqnarray}
Substituting Eqs. (\ref{pmetric}) and  (\ref{perttheta1}) into
Eq.(\ref{pertg}) and after tedious manipulations, we find the
perturbation equations for ten components as
\begin{eqnarray}
(t,t);&& e^{-ikt}E_1Y=0\nn\\
(t,r);&& e^{-ikt}E_2Y=0\nn\\
(t,\varphi_1);&&e^{-ikt}\Big(E_3
\partial_{\varphi_1}Y
+O_1\partial_{\varphi_2}Y\Big)=0\nn\\
(t,\varphi_2);&&e^{-ikt}\Big(E_3 \partial_{\varphi_2}Y
+O_2\partial_{\varphi_1}Y\Big)=0\nn\\
(r,r);&&e^{-ikt}E_4 Y=0\nn\\
(r,\varphi_1);&&e^{-ikt}\Big(E_5
\partial_{\varphi_1}Y
+O_3\partial_{\varphi_2}Y\Big)=0\nn\\
(r,\varphi_2);&&e^{-ikt}\Big(E_5
\partial_{\varphi_2}Y
+O_4\partial_{\varphi_1}Y\Big)=0\nn\\
(\varphi_1,\varphi_1);&&e^{-ikt}\Big(E_{6}Y
+E_{7}\partial_{\varphi_1}^2Y+O_{5}\partial_{\varphi_2} Y
+O_{6}\partial_{\varphi_1}\partial_{\varphi_2}Y\Big)=0\nn\\
(\varphi_1,\varphi_2);&&e^{-ikt}\Big(E_{8}\partial_{\varphi_2}Y
+E_{7}\partial_{\varphi_1}\partial_{\varphi_2}Y
+O_7Y+O_8\partial_{\varphi_1}^2Y\Big)=0\nn\\
(\varphi_2,\varphi_2);&&e^{-ikt}\Big(E_{9}Y
+E_{7}\partial_{\varphi_2}^2Y+E_{10}\partial_{\varphi_1}Y
+O_{9}\partial_{\varphi_2}Y
+O_{10}\partial_{\varphi_1}\partial_{\varphi_2}Y\Big)=0\label{comps},
\end{eqnarray}
where $E_{i}$ with $i=1,\cdots,10$ are functions of
($H_0,~H_1,~H_2,~K,~\Phi$) and $O_{i}$ with $i=1,\cdots,10$ are
functions of ($h_0,~h_1,~\psi$) (see Appendix for the details). It
is important to note that for $L>1$, the perturbation equations
(\ref{comps}) imply twenty conditions like
\begin{eqnarray}
E_{i}=0,~~~O_{i}=0,~~~ {\rm for}~i=1,\cdots,10
\end{eqnarray}
which mean that ten perturbation equations can be decoupled into two
classes:  odd-parity ($\{O_{i}\}$) and even-parity ($\{E_{i}\}$).

For the even-parity case, we observe that the condition of $E_7=0$
yields
\begin{equation}\label{const1}
H_0(r)-f^2H_2(r)-\frac{2f}{\bar{\phi}r}\Phi(r)=0.
\end{equation}
By using the above condition together with $E_{i}=0$
$(i=1,\cdots,6)$, one finds the central constraint equation as
\begin{eqnarray}\label{const2}
&&\hspace*{-2em}\left\{\lambda
f^{-1}-2+rf^{-1}f^{\prime}\right\}H_0+
\left\{2k^2r^2f^{-1}+2f+rf^{\prime}+\frac{r^2}{2}f^{-1}(f^{\prime})^2-\lambda\right\}K\nn\\
&&\hspace*{-2em}-\left\{2ikr+\frac{\lambda}{2ik}\right\}H_1-\left\{2\lambda-4f-2k^2r^2f^{-1}
-\frac{r^2}{2}f^{-1}(f^{\prime})^2\right\}\frac{\Phi}{\bar{\phi}r}=0,
\end{eqnarray}
where $\lambda=L(L+1)$. Manipulating  two equations of $E_2=0$ and
$E_3=0$ by using the Eqs.(\ref{const1}) and (\ref{const2}) lead to
\begin{eqnarray}
&&\frac{d}{dr}\left(K+\frac{\Phi}{\bar{\phi}r}\right) =\frac{\lambda
r f^{-1}f^{\prime}-4k^2r^2f^{-1}-6rf^{\prime}}
{2r(\lambda-2f+rf^{\prime})}\left(K+\frac{\Phi}{\bar{\phi}r}\right)+\frac{2\lambda
f-4k^2r^2-\lambda^2}
{2ir^2(\lambda-2f+rf^{\prime})}\left(\frac{H_1}{k}\right),\nn\\
&&\label{E2}\\
&&\hspace*{2em}\frac{d}{dr}\left(\frac{H_1}{k}\right)
=\frac{2\lambda-4f-2k^2r^2f^{-1}-r^2f^{-1}f^{\prime 2}/2}
{i(\lambda f-2f^2+rff^{\prime})}\left(K+\frac{\Phi}{\bar{\phi}r}\right)\nn\\
&&\hspace*{9em}-\frac{3\lambda
f^{-1}f^{\prime}/2-2f^{\prime}+rf^{-1}f^{\prime 2}-2k^2rf^{-1}}
{\lambda-2f+rf^{\prime}}\left(\frac{H_1}{k}\right).\label{E3}
\end{eqnarray}
Now we introduce the tortoise coordinate ($r^{*}=\int \frac{dr}{f}$)
and a new field defined by
\begin{equation}\label{calM}
\hat{\cal{M}}=\frac{1}{pq-h}\left\{p\left(K+\frac{\Phi}{\bar{\phi}r}\right)
-\frac{H_1}{k}\right\},
\end{equation}
where
\begin{eqnarray}
q(r)&=&\frac{\tilde{\lambda}(\tilde{\lambda}+1)r^2+3\tilde{\lambda}
Mr+6M^2}{r^2(\tilde{\lambda} r+3M)},~~ h(r)=\frac{i(-\tilde{\lambda}
r^2+3\tilde{\lambda}
Mr+3M^2)}{(r-2M)(\tilde{\lambda} r+3M)},\nn\\
~~~p(r)&=&-\frac{ir^2}{r-2M},~~ \tilde{\lambda}=\frac{\lambda}{2}-1.
\end{eqnarray}
As a result, from the Eqs.(\ref{E2}), (\ref{E3}) and (\ref{calM}) we
arrive at the Zerilli equation
\begin{eqnarray} \label{evenz}
\frac{d^2{\cal\hat{M}}}{dr^{*2}}+\Big[k^2-V_{Z}\Big]{\cal\hat{M}}=0,
\end{eqnarray}
where the Zerilli potential is given by~\cite{Zeri,CL}
\begin{equation}
V_{Z}(r)=f
\Bigg[\frac{2{\tilde{\lambda}}^2(\tilde{\lambda}+1)r^3+6\tilde{\lambda}^2Mr^2
+18\tilde{\lambda} M^2 r+18M^3} {r^3(\tilde{\lambda} r+3M)^2}\Bigg].
\end{equation}
The potential $V_{Z}(r^*)$ is always positive for whole range of
$-\infty \le r^* \le \infty$ , which implies that the even-parity
perturbation is stable, even though the scalaron $\Phi$ is coupled
to making the even-perturbation~\cite{MMS}.  In addition,  using the
tortoise coordinate ($r^{*})$, the scalaron equation (\ref{pertphi})
becomes
\begin{eqnarray}
\frac{d^2}{dr^{*2}}\Phi+\Big[k^2-V_{\Phi}\Big]\Phi&=&0,
\end{eqnarray}
where the scalaron potential $V_{\Phi}$ is given by
\begin{eqnarray}
V_{\Phi}=f\Big(\frac{\lambda}{r^2}+\frac{2M}{r^3} +m_{\phi}^2\Big)
\end{eqnarray}
with $m_{\phi}^2=\bar{\phi}V^{\prime\prime}(\bar{\phi})/3.$ The
potential $V_{\Phi}$ is always positive exterior the event horizon
if the mass squared $m_{\Phi}^2$ is positive~\cite{MMS}.\footnote{In
the original $f(R)$ gravity, the quantity of
$\bar{\phi}V^{\prime\prime}(\bar{\phi})/3$ corresponds to
$f^{\prime}(0)/3f^{\prime\prime}(0)$ [$\bar{\phi}\Leftrightarrow
f^{\prime}(0)$, $V^{\prime\prime}\Leftrightarrow
1/f^{\prime\prime}(0)]$. Therefore, the condition of  $m_{\Phi}^2>0$
implies  no-tachyon ($f^{\prime\prime}(0)>0$) if $f^{\prime}(0)>0$
(no-ghost) in $f(R)$ gravity.}

On the other hand,  for odd-parity perturbation ($\{O_{i}=0\}$),
the first five equations provide three:
\begin{eqnarray}
&& \underline{ O_1=0~{\rm or}~O_2=0}\nn\\
&&\hspace*{2em}r^3(-4M+\lambda r)h_0-r
f\Big(2kir^4-\frac{12}{\bar{\phi}}M\psi+kir^5h_1^{\prime}
+\frac{6}{\bar{\phi}}Mr\psi^{\prime}+r^5h_0^{\prime\prime}\Big)=0,
\label{feq1}\\
&&\underline{O_3=0~{\rm or}~O_4=0}\nn\\
&&\hspace*{2em}
-ikr^3\Big(2h_0-ikrh_1-rh_0^{\prime}\Big)+r^2f(\lambda-2)h_1+\frac{6}{\bar{\phi}}ikM\psi=0,\label{feq2}\\
&&\underline{O_5=0}\nn\\
&&\hspace*{2em}ikr^3h_0-(2M-r)\Big\{2Mh_1-(2M-r)rh_1^{\prime}\Big\}=0\label{feq3}
\end{eqnarray}
and all remaining equations $O_i$ with $i=6,\cdots,10$ are
redundant. Introducing the tortoise coordinate and a new field $Q$
defined by $Q=fh_1/r$, the above three equations
(\ref{feq1})$\sim$(\ref{feq3}) become one coupled second-order
equation
\begin{eqnarray}\label{Qeq}
\frac{d^2}{dr^{*2}}\tilde{Q}
+\Bigg\{k^2-f\Big(\frac{\lambda}{r^2}-\frac{6M}{r^3}\Big)\Bigg\}\tilde{Q}
&=&\frac{6ikMf}{r^5}\psi,
\end{eqnarray}
where $\tilde{Q}=\bar{\phi}Q$. Also, the perturbation equation
(\ref{perttheta}) for the dynamical scalar $\theta$ becomes a
coupled second-order equation
\begin{eqnarray}\label{psieq}
\frac{d^2}{dr^{*2}}\psi+\Bigg[k^2-f\Big\{\frac{\lambda}{r^2}
\Big(1+\frac{18M^2}{r^6\tilde{\alpha}}\Big)+\frac{2M}{r^3}\Big\}\Bigg]\psi&=&
-\frac{3\lambda(\lambda-2) i M f}{kr^5\tilde{\alpha}}\tilde{Q},
\end{eqnarray}
where $\tilde{\alpha}=\bar{\phi}\alpha$. This is an important
feature of CS coupling to $f(R)$ gravity. Actually these coupled
equation are the same found in~\cite{MPCG}. Hence, it is clear
that the black hole is stable against the perturbations of
$\tilde{Q}$ and $\psi$ when using two independent numerical
approaches of time evolution and a formation of frequency domain
employed in Ref.\cite{MPCG}.

Finally, we wish to mention the $f(R)$-form dependence on the
stability of the Schwarzschild black hole. In writing down two
Eqs.(\ref{Qeq}) and (\ref{psieq}), we introduce two new variables
$\tilde{Q}=\bar{\phi}Q$ and $\tilde{\alpha}=\bar{\phi} \alpha$
which show the connection to the original $f(R)$ gravity because
$\bar{\phi}=f'(0)$. As was mentioned in footnote 1, our analysis
is valid for a limited form of $f(R)=a_1R+a_2R^2+\cdots$. In this
limit from, we have $f'(0)= a_1=\bar{\phi}$, which is fixed by
choosing the limited $f(R)$ gravity. In general, we can say that
different $f(R)$ theories with different corresponding
$\bar{\phi}$  have different model parameters. However, as far as
the constant curvature-black hole stability is concerned, we
expect that Eqs.(\ref{Qeq}) and (\ref{psieq}) remain unchanged
except $\bar{\phi}$, leading to the stable Schwarzschild black
hole.

\section{Discussions}
In this work, we have performed the stability analysis of the
Schwarzschild black hole in $f(R)$ gravity with the parity-violating
CS term coupled to a dynamical scalar field $\theta$.  In order to
avoid the difficulty with fourth-order derivative terms appeared in
$f(R)$ gravity, we first transformed the $f(R)$ gravity into the
scalar-tensor theory by introducing a scalaron $\phi$. This will
provide the DCS modified gravity with two scalars, which provides
four physically propagating degrees of freedom.

Interestingly, the perturbation equation for the CS scalar $\theta$
is coupled to the odd-parity metric perturbation equation, providing
a system of two coupled second-order equations, while the scalaron
$\phi$ is coupled to the even-parity perturbation equation. This
enables us to perform the stability analysis of the Schwarzschild
black hole obtained from $f(R)$+DCS modified gravity theory
completely. It was shown  that the CS coupling affects the
Regge-Wheeler equation significantly, while $f(R)$ gravity does not
affect the Zerilli equation. It turns out that the Schwarzschild
black hole is stable against four external perturbations of
$\{\hat{\cal M},\Phi,\tilde{Q},\psi\}$ if the scalaron is free from
the tachyon.

However, the role of DCS term is limited here because its
perturbation $\delta C_{\mu\nu}$ in (\ref{cottonp2}) does not
involve third-order derivative terms. This higher derivative may
appear when the background solution contains a spherically symmetric
CS scalar~\cite{Yunes}. In this case, one could not decouple five
massive gravitons successfully because there exists a mixing between
odd- and even-parity modes,  and third-order derivative terms are
present. Even in the Minkowski background, it is not clear which
modes are propagating with their own  masses. It has been argued that a
spacelike vector of $v^\mu=\partial^\mu \bar{\theta}=(0,\vec{\mu})$
renders the theory free from ghosts and tachyons, while a timelike
vector of $v^\mu=(\mu,\vec{0})$ yields an inconsistent quantum
theory~\cite{BBBH,AK}. On the contrary, the opposite case is true:
the only tachyon- and ghost-free model is the one with a timelike
vector~\cite{PHH}. Hence, it seems to be a formidable task to
perform the stability of the Schwarzschild black hole when including
the third-order derivative terms.

 \vspace{1cm}

{\bf Acknowledgments}

This work was supported by the National Research Foundation of Korea
(NRF) grant funded by the Korea government (MEST) through the Center
for Quantum Spacetime (CQUeST) of Sogang University with grant
number 2005-0049409. Y. Myung  was partly  supported by the National
Research Foundation of Korea (NRF) grant funded by the Korea
government (MEST) (No.2011-0027293).

\newpage
\section*{Appendix: The explicit forms of twenty perturbation equations of $E_{i}=0$ and
$O_{i}=0$ where}
\begin{eqnarray}
&&
E_1=\frac{1}{2r^5}\Big[2r^5k^2K(r)+r(-2M^2f^{-1}+r^2\lambda)H_0(r)
+2i(3M-2r)r^3kH_1(r)\nn\\
&&\hspace*{3em}-rf(2M^2-r^4k^2)H_2(r)+2Mr^3fK^{\prime}(r)
+r^3(5M-2r)H_0^{\prime}(r)\nn
\nn\\&&\hspace*{3em}-2ir^5fkH_1^{\prime}(r)-Mr^3f^2H_2^{\prime}(r)
-r^5fH_0^{\prime\prime}(r)+\frac{1}{3}r^4f\bar{V}^{\prime\prime}\Phi(r)
\nn\\
&&\hspace*{3em}+\frac{2}{\bar{\phi}}\Big\{(2M-Mr+k^2r^4)\Phi(r)
+Mr^2f\Phi^{\prime}(r)\Big\}\Big],\nn\\
&& E_2=\frac{i}{2r^2}\Big[2(r-3M)k f^{-1}
K(r)-2krfH_2(r)+4kr^2K^{\prime}(r)-i\lambda H_1(r)\nn\\
&&\hspace*{13em}-\frac{2k}{\bar{\phi}}
\Big\{\left(1-\frac{M}{r}\right)f^{-1}\Phi(r)-r\Phi^{\prime}(r)\Big\}\Big],\nn\\
&& E_3=\frac{i}{2r^2}\Big[kr^2K(r)-2iMH_1(r)+kr^2fH_2(r)-ir^2
fH_1^{\prime}
+\frac{2kr}{\bar{\phi}}\Phi(r)\Big],\nn\\
&& E_4=\frac{1}{2r^4f^2}\Big[2M(2r-3M)f^{-1}H_0(r)+2iMkr^2H_1(r)
-f\Big(6M^2-4Mr\nn\\
&&\hspace*{3em}+k^2r^4-r^2f\lambda\Big)H_2(r)-2r(6M^2-7Mr+2r^2)K^{\prime}(r)
-Mr^2H_0^{\prime}
\nn\\
&&\hspace*{3em}+2ikr^4fH_1^{\prime}+r
f(6M^2-7Mr+2r^2)H_2^{\prime}-2r^4f^2K^{\prime\prime}(r)
+r^4fH_0^{\prime\prime}
\nn\\
&&\hspace*{1.7em}-\frac{r^3f}{3}\bar{V}^{\prime\prime}\Phi(r)
-\frac{2f}{\bar{\phi}}\Big\{(-5M+2r)\Phi(r)+r(5M-2r)\Phi^{\prime}(r)
+r^3f\Phi^{\prime\prime}(r)\Big\}\Big],\nn\\
 &&
E_5=\frac{-1}{2r^3f}\Big[(r-M)rf^{-1}H_0(r)-ikr^3H_1(r)
-(2M^2-3Mr+r^2)H_2(r)\nn\\
&&\hspace*{3em}+r^3fK^{\prime}(r)-r^3H_0^{\prime}
+\frac{2rf}{\bar{\phi}}\Big\{-2\Phi(r)+r\Phi^{\prime}(r)\Big\}\Big],\nn\\
&& E_6=\frac{-1}{2r^2}\Big[2Mrf^{-1}H_0(r)-2ikr^3H_1(r)+2
(2M^2+Mr-r^2)H_2(r)\nn\\
&&\hspace*{3em}+r^2(k^2r^2f^{-1}+\lambda+2)K(r)
-2r^2(3M-2r)K^{\prime}(r)-r^3H_0^{\prime}(r)
\nn\\
&&\hspace*{3em}-r^3f^2H_2^{\prime}(r)+r^4fK^{\prime\prime}(r)
+2rf\Big\{\frac{r^2}{6}f^{-1}\bar{V}^{\prime\prime}\Phi(r)
+\frac{1}{\bar{\phi}}\Big(r\Phi^{\prime}(r)-\Phi(r)\Big)\Big\}\Big],\nn\\
&& E_7=
\frac{1}{2r}\Big[rf^{-1}H_0(r)-rfH_2(r)-\frac{2}{\bar{\phi}}\Phi(r)\Big],\nn\\
&&
E_8=-E_7\cot\varphi_1,\nn\\
&&
E_{9}=E_{6}\sin^2\varphi_1,\nn\\
&& E_{10}=E_{7}\cos\varphi_1\sin\varphi_1,\nn
\end{eqnarray}
\begin{eqnarray}
&&\hspace*{-1.5em}
O_1=\frac{\csc\varphi_1}{2r^3}\Big[ir^2f\Big\{2kh_1(r)+krh_1^{\prime}(r)
-irh_0^{\prime\prime}(r)\Big\}+(4M-\lambda r)h_0(r)\nn\\
&&\hspace*{5em}-\frac{6}{\bar{\phi}r^2}Mf\Big\{2\psi(r)-r\psi^{\prime}(r)\Big\}\Big],\nn\\
&&\hspace*{-1.5em} O_2=\frac{-1}{2r^3}\Big[ir^2f\sin\varphi_1
\Big\{2kh_1(r)+krh_1^{\prime}(r)-irh_0^{\prime\prime}(r)\Big\}
-(rf+2M\cos2\varphi_1)\csc\varphi_1h_0(r)\nn\\
&&\hspace*{4em}+r(\csc\varphi_1-\lambda\sin\varphi_1)h_0(r)
-\frac{6}{\bar{\phi}r^2}Mf\sin\varphi_1\Big\{2\psi(r)-r\psi^{\prime}(r)\Big\}\Big],\nn\\
&&\hspace*{-1.5em}
O_3=\frac{\csc\varphi_1}{2r^3f}\Big[2ikr^2h_0(r)-ikr^3h_0^{\prime}(r)+
(2rf+k^2r^3-\lambda rf)h_1(r)-\frac{6i}{\bar{\phi}r}kM\psi(r)\Big],\nn\\
&&\hspace*{-1.5em}
O_4=\frac{-1}{2r^3f}\Big[ikr^2\sin\varphi_1\{2h_0(r)-rh_0^{\prime}\}
+\Big\{-rf\cos\varphi_1\cot\varphi_1+(rf+k^2r^3)\sin\varphi_1\nn\\
&&\hspace*{5em}+rf(\csc\varphi_1-\lambda\sin\varphi_1)\Big\}h_1(r)-\frac{6i}{\bar{\phi}r}kM\sin\varphi_1\psi(r)\Big],\nn\\
&&\hspace*{-1.5em}
O_5=\frac{\csc\varphi_1\cot\varphi_1}{r^3f}
\Big[ikr^3h_0(r)+rf\Big\{2Mh_1(r)+r^2fh_1^{\prime}(r)\Big\}\Big],\nn\\
&&\hspace*{-1.5em}
O_6=-O_5\tan\varphi_1,\nn\\
&&\hspace*{-1.5em}
O_7=\frac{ O_5 \lambda}{2}\Big[\sin^2\varphi_1\tan\varphi_1\Big],\nn\\
&&\hspace*{-1.5em}
O_8=O_5\sin^2\varphi_1\tan\varphi_1,\nn\\
&&\hspace*{-1.5em}
O_9=-O_5\sin^2\varphi_1,\nn\\
&&\hspace*{-1.5em} O_{10}=O_8.\nn
\end{eqnarray}

\end{document}